\documentclass[%
 reprint,
showpacs,preprintnumbers,
 amsmath,amssymb,
 aps,
 prd,
]{revtex4-1}

\usepackage{graphicx}
\usepackage{dcolumn}
\usepackage{bm}

\begin{document}

\title{Adiabatic and non-adiabatic perturbations for loop quantum cosmology}
\author{Yu Li}
\email{leeyu@mail.bnu.edu.cn}
  \affiliation{Department of Physics, Beijing Normal University, Beijing 100875, China}
\author{Jian-Yang Zhu}
\thanks{Author to whom correspondence should be addressed}
\email{zhujy@bnu.edu.cn}
\affiliation{Department of Physics, Beijing Normal University, Beijing 100875, China}
\date{\today}

\begin{abstract}
We generalize the perturbations theory of loop quantum cosmology to a hydrodynamical form and define an effective curvature perturbation on an uniform density hypersurfaces $\zeta_e$. As in the classical cosmology, $\zeta_e$ should be gauge-invariant and conservation on the large scales. The evolutions of both the adiabatic and the non-adiabatic perturbations for a multi-fluids model are investigated in the framework of the effective hydrodynamical theory of loop quantum cosmology with the inverse triad correction. We find that, different from the classical cosmology, the evolution of the large-scales non-adiabatic entropy perturbation can be driven by an adiabatic curvature perturbation and this adiabatic source for the non-adiabatic perturbation is a quantum effect. As an application of the related formalism, we study a decay model and give out the numerical results.

\end{abstract}


\maketitle


\section{\label{s1}Introduction}
The theory of the cosmological perturbations has been become a cornerstone of the modern cosmology.
It provides the key to understand the early evolution and the current large scale structure of our universe.
And, it is also used to describe the growth of the structure in the universe, calculate the predicted microwave background fluctuations, and in many other considerations. A widely used reference work about the cosmological perturbations can be seen in \cite{per1}.

Generally speaking, the origin of the cosmological perturbations is believed to come from the quantum fluctuation during the inflation. Therefore, it is interesting to study a possible quantum gravity effect in the cosmological perturbations theory. However, the problem of finding the quantum theory of the gravitational field is still open. One of the most active of the current approaches is loop quantum gravity. Loop quantum gravity (LQG) \cite{lqg1,lqg2,lqg3} is a mathematically well-defined, non-perturbative and background independent quantization of general relativity. Its cosmological version, the loop quantum cosmology (LQC) \cite{B1-B4} has achieved many successes. A major success of LQC is the resolution of the Big Bang singularity \cite{bb,nbb1,nbb2}; this result depends crucially on the discreteness of the spacetime geometry. With such a result, the big-bang singularity will be avoided through a big-bounce mechanism in the high energy region. In addition, LQC can also setup a suitable initial conditions for a successful inflation \cite{if1,if2} as well as possibly leaving an imprint in the cosmic microwave background \cite{if2}.

There are two types of the quantum corrections that are expected from the Hamiltonian of LQG. The one is called "the inverse triad correction" and the other is called "the holonomy correction" (see a review article \cite{LXZ}). The application of them on the scalar mode of perturbation can be found in \cite{lqcs}, the vector mode in \cite {lqcv} and the tensor mode in \cite{lqct}. The application of higher order holonomy corrections \cite{hhc} to the perturbations theory of cosmology is studied in \cite{hh2}.

Because the gauge-invariant approach for the cosmological perturbations with the holonomy correction has not been built up, in this paper we focus on the effective theory of LQC with the inverse triad correction. The gauge-invariant approach for the cosmological perturbations with the inverse triad correction has been built up in \cite{lqcg1,lqcg2,lqcg3}.

On the other hand, the perturbations considered in most of works are the adiabatic perturbation, for which the
density fluctuation is proportional to the pressure perturbation \cite{per2,per3,per4,per5}. However, there are some more detailed models such as the reheating at the end of inflation \cite{reheating}, or the effect of late-decaying scalar fields \cite{reh} require to discuss the multi-component system where it is useful to identify the gauge-invariant adiabatic and non-adiabatic modes.

In the classical cosmology, the most important result of the non-adiabatic perturbation is that \cite{nona1,nona2} the non-adiabatic (entropy) perturbation evolves independently of the curvature perturbation on the large scales, but that the evolution of the large-scale curvature perturbation is sourced by the non-adiabatic perturbation. In other words, the non-adiabatic perturbation can translate into the curvature perturbation on the large scales, while the curvature perturbation can not change to the entropy perturbation.

However, in loop quantum cosmology, the inverse triad correction will lead to a completely different evolution compared with the classical universe. Therefore, it is interesting to discuss the relationship between the adiabatic and the non-adiabatic perturbations under the theoretical framework of LQC.

The paper is organized as follows. At first, the gauge-invariant formalism of the cosmological perturbations theory with the inverse triad correction is reviewed briefly in Sec. \ref{s2}. Then in Sec. \ref{s3}, the perturbations theory of loop quantum cosmology is generalized to the hydrodynamical form. In Sec. \ref{s4}, the evolution of the non-adiabatic perturbation and the relationship between the adiabatic and the non-adiabatic perturbations under the theoretical framework of LQC are analyzed in detail. As an application of the related formalism, in Sec. \ref{s6} we study a decay model. The last Sec. \ref{s5} is the summary and conclusions.

\section{\label{s2}Scalar perturbation of LQC}
In this section, we review briefly the effective theory of LQC and the gauge-invariant formalism of the cosmological perturbations theory with the inverse triad correction. We give the background equations of LQC and the evolution equations of perturbation. In this paper, we only focus on the scalar mode perturbation which, along with the background FRW metric, takes the form
\begin{equation}\label{e1}
ds^2=a^2(\eta)[-(1+2\phi)d\eta^2+(1-2\psi)\delta_{ij}dx^idx^j],
\end{equation}
where the scale factor $a$ is a function of the conformal time $\eta$, the spatial indices $i$ and $j$ run from 1 to 3, and $\phi$ and $\psi$ are the scalar modes of the metric perturbation \cite{per5}.

\subsection{\label{s2a}Background}

The universe considered in this paper is fulfilled by a scalar field $\varphi$ with the potential $V$, so the effective Friedmann equation with the inverse triad correction of LQC can be written as \cite{lqcs}
\begin{equation}\label{e2}
{\cal{H}}^2=\frac{8\pi G}{3}\alpha\left[\frac{\varphi'^2}{2\nu}+pV(\varphi)\right],
\end{equation}
where $G$ is the gravitational constant, $p=a^2$ the triad variable in minisuperspace \cite{nbb1},
${\cal{H}}\equiv a'/a=p'/2p$ the Hubble parameter; the prime "$'$" denotes the derivative with respect to the conformal time, and $\alpha$ and $\nu$ are the parameters characterizing the effective inverse triad correction \cite{lqcg3}.
\begin{eqnarray}
\alpha&\approx& 1+\alpha_0 \delta_{pl},\label{e3}\\
\nu&\approx& 1+\nu_0 \delta_{pl},\label{e4}
\end{eqnarray}
where
\begin{equation}\label{e5}
\delta_{pl}\equiv\left(\frac{p_{pl}}{p}\right)^{\frac{\sigma}{2}}=\left(\frac{a_{pl}}{a}\right)^{\sigma},
\end{equation}
$p_{pl}=a_{pl}^2$ is constant. $\sigma$ is an ambiguity parameter for quantization, it depends on which the geometrical minisuperspace variable has an equi-distant stepsize in the dynamics. A detailed calculation then shows that the constant coefficients $\alpha_0$ and $\nu_0$ are \cite{anc}
\begin{eqnarray}\label{e9}
\alpha_0&=&\frac{(3q-\sigma)(6q-\sigma)}{2^23^4}\left(\frac{\Delta_{pl}}{p_{pl}}\right)^2,\\
\nu_0&=&\frac{\sigma(2-l)}{54}\left(\frac{\Delta_{pl}}{p_{pl}}\right)^2,
\end{eqnarray}
where $\Delta_{pl}\equiv2\sqrt{3}\pi\gamma\ell_{pl}^2$ is the area gap, and $\gamma$ is the Barbero-Immirzi parameter.
$1/2\leq l<1$ and $1/3\leq q<2/3$ are another sets of ambiguity parameters, they relate to different ways of quantizing the classical Hamiltonian.

From the definition of the Hubble parameter, it can be easily to check the relationship between the derivative with respect to the conformal time and the derivative with respect to $p$
\begin{equation}\label{e6}
'=2{\cal{H}}\frac{d}{d\ln p},
\end{equation}
therefore
\begin{equation}\label{e7}
\delta'_{pl}=-\sigma{\cal{H}}\delta_{pl}.
\end{equation}
And the effective Klein-Gordon equation can be read \cite{lqcg3}
\begin{equation}\label{e8}
\varphi''+2{\cal{H}}\left(1-\frac{d\ln \nu}{d\ln p}\right)\varphi'+\nu p V,_{\varphi}=0,
\end{equation}
where $V,_{\varphi}$ means $\partial V/\partial\varphi$.

\subsection{\label{s2b}Gauge-invariant formalism}
If one only consider the primary correction functions of $\alpha$ and $\nu$, there will be anomalies in the effective constraint algebra \cite{lqcs,lqcv,lqct}. To keep the consistency of the theory, one must introduce the counterterms which proportional to $\delta_{pl}$ \cite{lqcg1,lqcg2}. Explicitly, these counterterms are \cite{lqcg3}:
\begin{eqnarray}
f~&=&\frac{1}{\sigma}\frac{d\ln\alpha}{d\ln p}=-\frac{\alpha_0}{2}\delta_{pl},\label{e10}\\
f_1&=&f-\frac{1}{3}\frac{d\ln\nu}{d\ln p}=\frac{1}{2}\left(\frac{\sigma\nu_0}{3}-\alpha_0\right)\delta_{pl},\label{e11}\\
h~&=&2\frac{d\ln\alpha}{d\ln p}-f=\alpha_0\left(\frac{1}{2}-\sigma\right)\delta_{pl},\label{e12}\\
g_1&=&\frac{1}{3}\frac{d\ln\alpha}{d\ln p}-\frac{d\ln\nu}{d\ln p}+\frac{2}{9}\frac{d^2\ln \nu}{d\ln p^2}\nonumber\\
&=&\frac{\sigma}{2}\left(\frac{\sigma\nu_0}{9}+\nu_0-\frac{\alpha_0}{3}\right)\delta_{pl},\label{e13}\\
f_3&=&f_1-g_1\nonumber\\
&=&\frac{1}{2}\left[\alpha_0\left(\frac{\sigma}{3}-1\right)-\frac{2\sigma\nu_0}{3}\left(\frac{\sigma}{6}+1\right)\right]\delta_{pl}.\label{e14}
\end{eqnarray}
In this paper, we only focus on the linear order of counterterms as in \cite{lqcg3}, for instance, $(1+f)(1+h)=1+f+h+O(\delta_{pl}^2)$. We neglect the terms of order $O(\delta_{pl}^2)$.

If we ignore the anisotropy of the universe, two metric perturbations are proportional to each other \cite{lqcg2}
\begin{equation}\label{e15}
\phi=(1+h)\psi.
\end{equation}
One should note that, different from the classical cosmology, there is a counterterm $h$ in Eq.(\ref{e5}).

The effective dynamics equation for the metric perturbation is \cite{lqcg2}
\begin{widetext}
\begin{equation}\label{e16}
3{\cal{H}}(1+f)[\psi'+(1+f){\cal{H}}\phi]-\alpha^2\nabla^2\psi
=-4\pi G\frac{\alpha}{\nu}(1+f_3)[\varphi'\delta\varphi'-\varphi'^2(1+f_1)\phi
+\nu p V,_{\varphi}\delta\varphi],
\end{equation}
\end{widetext}
where $\delta\varphi$ is the perturbation of $\varphi$, and the effective Klein-Gordon equation for $\delta\varphi$ is \cite{lqcg3}
\begin{eqnarray}
\delta\varphi''+2{\cal{H}}B_1\delta\varphi'&-&(s^2\nabla^2-\nu p V,_{\varphi\varphi})\delta\varphi\nonumber\\
&-&B_2\varphi'\psi'+2B_3{\cal{H}}\varphi'\psi=0,\label{e17}
\end{eqnarray}
where
\begin{eqnarray}
B_1&=&1-\frac{d\ln \nu}{d\ln p}-\frac{d g_1}{d\ln p}=1+B_{10}\delta_{pl},\label{e18}\\
B_2&=&4+f_1+h+3g_1=4+B_{20}\delta_{pl},\label{e19}\\
B_3&=&(1+f_1+h)\frac{\nu p V,_{\varphi}}{{\cal{H}}\varphi'}-\frac{dh}{d\ln p}-\frac{df_3}{d\ln p}\nonumber\\
&=&-2-\frac{\varphi''}{{\cal{H}}\varphi'}+B_{30}\delta_{pl},\label{e20}
\end{eqnarray}
and
\begin{eqnarray}
B_{10}&\equiv&\sigma\left[\nu_0\left(\frac{\sigma}{6}+1\right)-\frac{\alpha_0}{2}\right],\label{e21}\\
B_{20}&\equiv&\frac{\sigma}{2}\left(\frac{\sigma\nu_0}{3}+\frac{10\nu_0}{3}-3\alpha_0\right),\label{e22}\\
B_{30}&\equiv&\sigma\left[\left(\frac{\nu_0}{6}-\alpha_0\right)(1-\frac{\varphi''}{{\cal{H}}\varphi'})\right.\nonumber\\
&&\left.-\nu_0\left(\frac{\sigma}{12}+2\right)+\frac{\alpha_0}{2}(7-\sigma)\right],\label{e23}
\end{eqnarray}
and $s^2=\alpha^2(1-f_3)$ is the squared propagation speed of the perturbation.

The other dynamics equations can be seen in \cite{lqcg2}.

\section{\label{s3}Effective hydrodynamical perturbation}

One can notice that, the dynamics equations of the perturbation Eq.(\ref{e16}) and Eq.(\ref{e17}) are all based on the matter of scalar field. For discussing the multi-components model, it is convenient to generalize the model to a hydrodynamical form, in which the universe is fulfilled by a general fluid.

However, the theory of the cosmological perturbations with the inverse-triad corrections based on the general fluid model has not been built up. This is what we do in this section.

We adopt a simple strategy. Firstly, we rewrite the effective equations to a hydrodynamical form, then we define an effective density $\rho_e$, an effective pressure $P_e$, and their perturbations $\delta\rho_e$ and $\delta P_e$. And we represent the evolution equations by these effective quantity.
Secondly, we assume that, the hydrodynamical form of these equations with the inverse-triad corrections of LQC is also can be applied to the general fluid models.

Therefore, there are two requirements for our hydrodynamical form.
\begin{itemize}
\item The classical limit of the effective density, the effective pressure and their perturbations should coincide with the classical density and pressure of fluid.
\item The classical limit of the effective hydrodynamical perturbation equations with inverse-triad corrections of LQC should coincide with the classical equations \cite{per1}.
\end{itemize}

\subsection{\label{s3a}Effective hydrodynamical equations}
At first, we give the background equations in fluid form. The effective Friedmann equation in fluid form can be seen in \cite{lqcg1}:
\begin{equation}\label{e24}
{\cal{H}}^2=\frac{8\pi G}{3}(\alpha p) \rho_e.
\end{equation}
Compared with Eq.(\ref{e2}), we have $\rho_e\equiv\frac{\varphi'^2}{2p\nu}+V(\varphi)$, where the subscript "e" means the "effective".

From Eq (\ref{e8}) and Eq.(\ref{e24}), one can obtain the effective continuity equation:
\begin{equation}\label{e25}
\rho_e'+3{\cal{H}}\left(1-\frac{1}{3}\frac{d \ln \nu}{d \ln p}\right)(\rho_e+P_e)=0,
\end{equation}
where $P_e\equiv\frac{\varphi'^2}{2p\nu}-V(\varphi)$ is the effective pressure.

Equation (\ref{e16}) inspires us to define an effective density perturbation as follow:
\begin{equation}\label{e26}
\delta\rho_e\equiv\frac{1}{p\nu}[\varphi'\delta\varphi'-(1+f_1)\varphi'^2\phi]+V,_{\varphi}\delta\varphi,
\end{equation}
and we define an effective pressure perturbation analogously:
\begin{equation}\label{e27}
\delta P_e\equiv\frac{1}{p\nu}[\varphi'\delta\varphi'-(1+f_1)\varphi'^2\phi]-V,_{\varphi}\delta\varphi.
\end{equation}
Under these definitions, Eq.(\ref{e16}) changes to
\begin{eqnarray}\label{e28}
3{\cal{H}}(1+f)[\psi'+(1+f){\cal{H}}\phi]&-&\alpha^2\nabla^2\psi\nonumber\\
=-4\pi G\alpha p(1&+&f_3)\delta\rho_e.
\end{eqnarray}
From Eq(\ref{e8}), Eq.(\ref{e17}) and Eq.(\ref{e26}), it can be verified that $\delta\rho_e$ is suitable for equation as follow
\begin{widetext}
\begin{equation}\label{e29}
\delta\rho_e'+3{\cal{H}}\left(1-\frac{1}{3}\frac{d\ln \nu}{d\ln p}-\frac{1}{3}\frac{dg_1}{d\ln p}\right)(\delta\rho_e+\delta P_e)
-3(1+g_1)(\rho_e+P_e)\psi'-s^2\frac{\nabla^2\varphi'\delta\varphi}{p\nu}=0.
\end{equation}
\end{widetext}
On the large scales limit, where the $\nabla^2$ term tends to vanish, Eq.(\ref{e29}) changes to
\begin{equation}\label{e30}
\delta\rho_e'+3{\cal{H}}D_1(\delta\rho_e+\delta P_e)-3D_2(\rho_e+P_e)\psi'=0,
\end{equation}
where
\begin{eqnarray}
D_1&\equiv&1-\frac{1}{3}\frac{d\ln \nu}{d\ln p}-\frac{1}{3}\frac{dg_1}{d\ln p},\\
D_2&\equiv&1+g_1.
\end{eqnarray}
At this point, the definitions of $\rho_e$, $P_e$, $\delta\rho_e$ and $\delta P_e$ as well as Eq.(\ref{e28}) and Eq.(\ref{e30})
satisfy the two requirements on the large scales above-mentioned.

At least on the large scales, we can rewrite the theory of the gauge-invariant perturbation of LQC to the hydrodynamical form. From now on, we assume that, these hydrodynamical equations are valid not only for the scalar field but also for a general fluid, and we restrict our discussion on the large scales limit.

\subsection{\label{s3b}Curvature perturbation on uniform density hypersurfaces}

It is convenient for the cosmological applications to introduce a curvature perturbation on an uniform density hypersurfaces $\zeta$, which first introduced by Bardeen, Steinhardt and Turner \cite{cp1} as a conserved quantity for the adiabatic perturbation on the large scales \cite{cp2}.

In the classical cosmology, the definition of $\zeta$ is
\begin{equation}\label{e31}
-\zeta\equiv\psi+{\cal{H}}\frac{\delta\rho}{\rho'}.
\end{equation}
We introduce a LQC correction $D_3$ in this definition
\begin{equation}\label{e32}
-\zeta_e\equiv\psi+D_3{\cal{H}}\frac{\delta\rho_e}{\rho_e'}.
\end{equation}
But we do not fix the form of $D_3$ right now.

For the classical cosmology, $\zeta$ is a conservative and gauge-invariant quantity on the large scales. Therefore, it is natural to require the effective curvature perturbation $\zeta_e$ is also has the similar properties.

The evolution equation of $\zeta_e$ can be obtained by taking the time derivative of Eq.(\ref{e32})
\begin{widetext}
\begin{equation}\label{e33}
\zeta_e'=-\psi'-\left[D_3'{\cal{H}}\frac{\delta\rho_e}{\rho_e'}+D_3\left({\cal{H}}'\frac{\delta\rho_e}{\rho_e'}+{\cal{H}}\frac{\delta\rho_e'}{\rho_e'}
-{\cal{H}}\delta\rho_e\frac{\rho_e''}{\rho_e'^2}\right)\right].
\end{equation}
From Eq.(\ref{e25}) we have
\begin{equation}\label{e34}
\rho_e''+3[{\cal{H}}'(1-f+f_1)+{\cal{H}}(f_1-f)'](\rho_e+P_e)
+3{\cal{H}}(1-f+f_1)(\rho_e'+P_e')=0.
\end{equation}
Substituting Eq.(\ref{e34}) to Eq.(\ref{e33}) we have
\begin{eqnarray}\label{e35}
\zeta_e'=-\psi'-\frac{1}{\rho_e'}\left\{D_3'{\cal{H}}\delta\rho_e+D_3{\cal{H}}'\delta\rho_e+D_3{\cal{H}}\delta\rho_e'
+3D_3{\cal{H}}\delta\rho_e\left[\frac{{\cal{H}}'}{\rho_e'}(1-f+f_1)(\rho_e+P_e)\right.\right.\nonumber\\
\left.\left.+\frac{{\cal{H}}}{\rho_e'}(f_1-f)'(\rho_e+P_e)+(1-f+f_1){\cal{H}}(1+c_{se}^2)\right]\right\},
\end{eqnarray}
where $c_{se}^2\equiv P_e'/\rho_e'$ is an effective adiabatic sound speed.
By using Eq.(\ref{e30}) and Eq.(\ref{e25}) we have
\begin{eqnarray}\label{e36}
\zeta_e'=-\frac{1}{\rho_e'}\left\{3{\cal{H}}(\rho_e+P_e)\psi'[D_2D_3-(1-f+f_1)]+\left[D_3'-\frac{D_3(f_1-f)'}{1-f+f_1}\right]{\cal{H}}\delta\rho_e\right.\nonumber\\
\left.+3D_3{\cal{H}}^2(\delta\rho_e+\delta P_e)(1-f+f_1-D_1)-3{\cal{H}}^2\delta P_{nad}\right\},
\end{eqnarray}
\end{widetext}
where $\delta P_{nad}\equiv\delta P_e-c_{se}^2\delta\rho_e$ is a non-adiabatic pressure perturbation.

One can notice that if we set $D_3=1-f+f_1$ and choose suitable $\sigma, \ell$ and $q$ to make $g_1=0$ ($D_2=1$ and $D_1=D_3$), we will find a simple relationship
\begin{equation}\label{e37}
\zeta_e'=-\frac{{\cal{H}}\delta P_{nad}}{(1-f+f_1)(\rho_e+P_e)}.
\end{equation}
So, the definition of $\zeta_e$ could be
\begin{equation}\label{e38}
-\zeta_e\equiv\psi+(1-f+f_1){\cal{H}}\frac{\delta\rho_e}{\rho_e'}.
\end{equation}
Under this definition, $\zeta_e$ is a conserved quantity on the large scales when we neglect the non-adiabatic perturbation (when $g_1=0$). This is the same as the classical cosmology. However, this definition of $\zeta_e$ is not necessarily gauge-invariant. We discuss this topic next.

Back to the scalar model, the gauge transformations for $\delta\varphi$, $\psi$ and $\phi$ are \cite{lqcg2}
\begin{eqnarray}
\delta\varphi&\rightarrow&\delta\varphi+\varphi'(1+f_1)\xi^0,\label{e39}\\
\psi&\rightarrow&\psi-{\cal{H}}(1+f)\xi^0,\label{e40}\\
\phi&\rightarrow&\phi+{\cal{H}}\xi^0+(\xi^{0})',\label{e41}
\end{eqnarray}
where $\xi^0$ is the 0-component of the infinitesimal coordinate transformations $\xi^\mu$.

From the definition Eq.(\ref{e26}) of $\delta\rho_e$, we can obtain both the gauge transformation for $\delta\rho_e$ and the gauge transformation for $\zeta_e$ as, respectively
\begin{eqnarray}\label{e42}
\delta\rho_e\rightarrow&&\delta\rho_e\nonumber\\
&&+\frac{\varphi'^2{\cal{H}}\xi^0}{p\nu}\left[\frac{f_1'}{{\cal{H}}}-(1+f_1)\left(3-2\frac{d\ln \nu}{d\ln p}\right)\right],
\end{eqnarray}
and
\begin{eqnarray}\label{e43}
\zeta_e&\rightarrow&\zeta_e+{\cal{H}}\xi^0(1+f)\nonumber\\
&&+{\cal{H}}\xi^0(1-f+f_1)\frac{\frac{f_1'}{{\cal{H}}}-(1+f_1)\left(3-2\frac{d\ln \nu}{d\ln p}\right)}{3-\frac{d\ln \nu}{d\ln p}}.
\end{eqnarray}
If we require $\zeta_e$ is gauge-invariant, we need
\begin{equation}\label{e44}
(1+f)+(1-f+f_1)\frac{\frac{f_1'}{{\cal{H}}}-(1+f_1)\left(3-2\frac{d\ln \nu}{d\ln p}\right)}{3-\frac{d\ln \nu}{d\ln p}}=0.
\end{equation}
From Eqs.(\ref{e10})-(\ref{e14}), then Eq.(\ref{e44}) can be reduced to
\begin{equation}\label{e45}
9\nu_0+\sigma\nu_0=3\alpha_0.
\end{equation}
This condition is the same as $g_1=0$. So, when $g_1=0$, $\zeta_e$, which we have defined in this paper, is a gauge-invariant and a conserved quantity on the large scales.

\section{\label{s4}Non-adiabatic perturbations}

A general thermodynamic system can be fully described by three variables: ($\rho, P, S$), where $\rho$ is the energy density, $P$ the pressure and $S$ the entropy. However, only two of them are independent. If we choose $\rho$ and $S$ as two independent variables, the pressure can be expressed as $P\equiv P(S,\rho)$. Then, the pressure perturbation can be expanded into a Taylor series as
\begin{equation}{\label{e68}}
\delta P=\frac{\partial P}{\partial S}\delta S+\frac{\partial P}{\partial \rho}\delta \rho.
\end{equation}
This can be recast in a more familiar form:
\begin{equation}{\label{e69}}
\delta P=\delta P_{nad}+c^2_s\delta\rho,
\end{equation}
where $c_s^2\equiv \left. \frac{\partial P}{\partial \rho }\right| _S$ is the adiabatic sound speed. If the system is adiabatic, which means $\delta S=0$, we can find that $\delta P=c^2_s\delta\rho$. Therefore, $c^2_s\delta\rho$ is the adiabatic part of $\delta P$ and $\delta P_{nad}\equiv \left. \frac{\partial P}{\partial S}\right| _{\rho} \delta S$ is the non-adiabatic part of it \cite{nona2}.

One can notice that, when the system is adiabatic, $\delta P=\frac{\partial P}{\partial \rho}\delta \rho$ means that $P$ is only a function of $\rho$, i.e. $P=P(\rho)$. Therefore, $P=P(\rho)$ can be seen as the adiabatic condition. By extension, for a thermodynamic quantity $\mathcal{G}$, its adiabatic condition is that it is only the function of $\rho$, i.e. $\mathcal{G}=\mathcal{G}(\rho)$.

\subsection{\label{s4a}Interacting fluids}

In this section, we will consider a multi-fluids model. The number of the interacting fluids in universe are arbitrary. Each fluid has an energy-momentum tensor $T^{\mu\nu}_{(\alpha)}$, the symbol $\alpha$ here denotes a different fluid, $\mu$ or $\nu$ denotes a space-time index, the inverse triad correction of LQC is not included. The total energy momentum tensor $T^{\mu\nu}=\sum_{\alpha}T^{\mu\nu}_{(\alpha)}$, is covariantly conserved, but, for energy, we allow transfer between the fluids.
\begin{equation}\label{e46}
\nabla_{\mu}T^{\mu\nu}_{(\alpha)}=Q_{\alpha}^{\nu},
\end{equation}
where $Q_{\alpha}^{\nu}$ is the quantity of the energy transferring in $\alpha$ fluid. The total energy-momentum tensor is conservation, i.e. $\nabla_{\mu}T^{\mu\nu}=0$, so it requires that $\sum_{\alpha}Q^{\nu}_{\alpha}=0$.

Total energy density and total pressure are
\begin{equation}\label{e47}
\rho_e=\sum_{\alpha}\rho_{e(\alpha)},\ \ \  P_e=\sum_{\alpha}P_{e(\alpha)}.
\end{equation}
The continuity equation for each individual fluid is thus \cite{per6}
\begin{equation}\label{e48}
\rho_{e(\alpha)}'+3{\cal{H}}\left(1-\frac{1}{3}\frac{d \ln \nu}{d \ln p}\right)(\rho_{e(\alpha)}+P_{e(\alpha)})=Q_{\alpha},
\end{equation}
where $Q_{\alpha}$ is the time component of the energy transferring vector $Q_{\alpha}^0$ and we assume the
inverse triad correction for all fluids are the same.

The perturbation for energy transferring can be written as $Q_{\alpha}\phi+\delta Q_{\alpha}$ \cite{per6}. So the evolution equation of the density perturbation for each individual fluid is
\begin{eqnarray}\label{e49}
\delta\rho_{e(\alpha)}'&+&3{\cal{H}}D_1(\delta\rho_{e(\alpha)}+\delta P_{e(\alpha)})\nonumber\\
&-&3D_2(\rho_{e(\alpha)}+P_{e(\alpha)})\psi'=Q_{\alpha}\phi+\delta Q_{\alpha}.
\end{eqnarray}

Analogous to the definition of $\zeta_e$, we can define the curvature perturbation for each individual fluid
\begin{equation}\label{e50}
-\zeta_{e(\alpha)}\equiv\psi+(1-f+f_1){\cal{H}}\frac{\delta\rho_{e(\alpha)}}{\rho_{e(\alpha)}'}.
\end{equation}
It can be  proved easily that the total curvature perturbation $\zeta_e$ is a weighted sum of the individual perturbation
\begin{equation}\label{e51}
\zeta_e=\sum_{\alpha}\frac{\rho_{e(\alpha)}'}{\rho_e'}\zeta_{e(\alpha)}.
\end{equation}
The difference between any two curvature perturbations describes a relative entropy (or isocurvature) perturbation
\begin{eqnarray}\label{e52}
{\cal{S}}_{\alpha\beta}&=&3(\zeta_{e(\alpha)}-\zeta_{e(\beta)})\nonumber\\
&=&-3(1-f+f_1){\cal{H}}\left(\frac{\delta\rho_{e(\alpha)}}{\rho_{e(\alpha)}'}-\frac{\delta\rho_{e(\beta)}}{\rho_{e(\beta)}'}\right).
\end{eqnarray}

\subsection{\label{s4b}Evolution equations}
In the multi-fluids model, the total non-adiabatic pressure perturbation $\delta P_{nad}$ may be split into two parts:
\begin{equation}\label{e53}
\delta P_{nad}\equiv\delta P_{intr}+\delta P_{rel},
\end{equation}
where $\delta P_{intr}\equiv\sum_{\alpha}\delta P_{intr(\alpha)}$, and $\delta P_{intr(\alpha)}$ is the intrinsic non-adiabatic pressure perturbation of each fluid, its definition is
\begin{equation}\label{e54}
\delta P_{intr(\alpha)}\equiv\delta P_{e(\alpha)}-c_{e\alpha}^2\delta\rho_{e(\alpha)},
\end{equation}
where $c_{e\alpha}^2\equiv P_{e(\alpha)}'/\rho_{e(\alpha)}'$ is the effective sound speed of each fluid. It is related to $c_{se}^2$ by
\begin{equation}\label{e55}
c_{se}^2=\sum_{\alpha}\frac{\rho_{e(\alpha)}'}{\rho_e}c_{e\alpha}^2.
\end{equation}

From the Eq.(\ref{e53}), one can notice that, the first part of $\delta P_{nad}$ comes from the intrinsic non-adiabatic perturbation of each individual fluid. And the other part $\delta P_{rel}$ should come from the relative entropy perturbation ${\cal{S}}_{\alpha\beta}$. From Eqs.(\ref{e52}) and (\ref{e53}) one can represent the $\delta P_{rel}$ by ${\cal{S}}_{\alpha\beta}$
\begin{equation}\label{e56}
\delta P_{rel}=-\frac{1+f-f_1}{6{\cal{H}}\rho_e'}\sum_{\alpha,\beta}\rho_{e(\alpha)}'\rho_{e(\beta)}'(c_{e\alpha}^2-c_{e\beta}^2){\cal{S}}_{\alpha\beta}.
\end{equation}
If $P_{e(\alpha)}=P_{e(\alpha)}(\rho_{e(\alpha)})$, the intrinsic non-adiabatic pressure perturbation
$\delta P_{intr(\alpha)}$ and $\delta P_{intr}$will vanish. Even so, the $\delta P_{rel}$ will not be vanished, because it depending on the curvature perturbation of individual fluid $\zeta_{e(\alpha)}$.

The evolution of $\zeta_{e(\alpha)}$ can be obtained from Eq.(\ref{e50})
\begin{eqnarray}
\zeta _{e(\alpha )}^{\prime } &=&\frac{3{\cal {H}}^2(1-f+f_1)^2\delta
P_{intr(\alpha )}}{\rho _{e(\alpha )}^{\prime }}  \nonumber \\
&&-(1-f+f_1)\frac{{\cal {H}}Q_\alpha }{\rho _{e(\alpha )}^{\prime }}\left\{
\frac{\psi ^{\prime }}{(1-f+f_1){\cal {H}}}\right.   \nonumber \\
&&+\frac{\delta \rho _{e(\alpha )}}{\rho _{e(\alpha )}^{\prime }}\left[ -%
\frac{Q_\alpha ^{\prime }}{Q_\alpha }+\frac{{\cal {H}}^{\prime }}{{\cal {H}}}%
+\frac{(f_1-f)^{\prime }}{1-f+f_1}\right]   \nonumber \\
&&\left. +\phi +\frac{\delta Q_\alpha }{Q_\alpha }\right\} .  \label{e57}
\end{eqnarray}
Similar in \cite{nona2}, we define an effective non-adiabatic perturbation of energy transferring by
\begin{eqnarray}\label{e58}
\delta Q_{nad(\alpha)}\equiv Q_{\alpha}\left\{
\frac{\psi'}{(1-f+f_1){\cal{H}}}
+\frac{\delta\rho_{e(\alpha)}}{\rho_{e(\alpha)}'}
\left[-\frac{Q_{\alpha}'}{Q_{\alpha}}\right.\right.\nonumber\\
\left.\left.+\frac{{\cal{H}}'}{{\cal{H}}}+\frac{(f_1-f)'}{1-f+f_1}\right]+\phi+\frac{\delta Q_{\alpha}}{Q_{\alpha}}\right\}.
\end{eqnarray}
Different from the classical situation, there is an effective quantum term in the non-adiabatic perturbation of energy transferring. It means that LQC quantum corrections will affect the energy transferring between different fluids. This influence is non-adiabatic. If there is no energy transferring between the different fluids, i.e. $Q_{\alpha}=0$, then $\delta Q_{nad(\alpha)}=0$. If $\delta P_{intr(\alpha)}$ vanishes also, the $\zeta_{e(\alpha)}$ will conserve on the large scales.
However, in general speaking  $\delta Q_{nad(\alpha)}\neq0$, therefore $\zeta_{e(\alpha)}$ is not conservation on the large scales.

There are two source for $\zeta_{e(\alpha)}$, one is an intrinsic non-adiabatic perturbation and the other is a non-adiabatic perturbation of energy transferring. The non-adiabatic perturbation of energy transferring is can also be split into two parts as the same of $\delta P_{nad}$
\begin{equation}\label{e59}
\delta Q_{nad(\alpha)}=\delta Q_{intr(\alpha)}+\delta Q_{rel(\alpha)},
\end{equation}
where the definition of the intrinsic part is
\begin{equation}
\delta Q_{intr(\alpha)}\equiv\delta Q_{\alpha}-(Q_{\alpha}'/\rho_{e(\alpha)}')\delta\rho_{e(\alpha)}.
\end{equation}
If $Q_{\alpha}=Q_{\alpha}(\rho_{e(\alpha)})$, then $\delta Q_{intr(\alpha)}=0$.
Thus, by using Eq.(\ref{e28}) and Eq.(\ref{e58}), one can obtain (notice that $g_1=0 \Rightarrow f_1=f_3$)
\begin{equation}\label{e60}
\delta Q_{rel(\alpha)}=\frac{Q_{\alpha}\rho_e'}{2\rho_e}\left(\frac{\delta\rho_{e(\alpha)}}{\rho_{e(\alpha)}'}-\frac{\delta\rho_e}{\rho_e'}\right)
+\frac{Q_{\alpha}{\cal{H}}\delta\rho_{e(\alpha)}}{\rho_{e(\alpha)}'}+Q_{\alpha}\mathcal{E}_{\alpha},
\end{equation}
where
\begin{equation}\label{e61}
\mathcal{E}_{\alpha}\equiv (2f-f_1)\left[(\sigma+1){\cal{H}}\frac{\delta\rho_{e(\alpha)}}{\rho_{e(\alpha)}'}+\zeta_{e(\alpha)}\right].
\end{equation}
$\mathcal{E}_{\alpha}$ is an effective quantum term which is vanish only in the classical limit. One should notice that it depends on the curvature perturbation of the individual fluid $\zeta_{e(\alpha)}$.

The first term of Eq.(\ref{e60}) can be  represented by ${\cal{S}}_{\alpha\beta}$
\begin{equation}\label{e62}
\frac{Q_{\alpha}\rho_e'}{2\rho_e}\left(\frac{\delta\rho_{e(\alpha)}}{\rho_{e(\alpha)}'}-\frac{\delta\rho_e}{\rho_e'}\right)
=-\frac{(1+f-f_1)Q_{\alpha}}{6{\cal{H}}\rho_e}\sum_{\beta}\rho_{e(\beta)}'{\cal{S}}_{\alpha\beta}.
\end{equation}
If we represent our discussion by cosmic time, then the second term of Eq.(\ref{e60}) can be absorbed into the first term \cite{nona2}. From Eq.(\ref{e56}), Eq.(\ref{e62}) and Eq.(\ref{e60}), we have
\begin{equation}\label{e63}
\delta P_{rel}=\frac{2\rho_e}{\rho_e'}\sum_{\alpha}\rho_{e(\alpha)}'c_{e(\alpha)}^2\left(\frac{\delta Q_{rel(\alpha)}}{Q_{\alpha}}
-\frac{{\cal{H}}\delta\rho_{e(\alpha)}}{\rho_{e(\alpha)}'}+\mathcal{E}_{\alpha}\right).
\end{equation}
We find that there is also an effective quantum term in $\delta P_{rel}$.

From the definition of ${\cal{S}}_{\alpha\beta}$, we know that
\begin{equation}\label{e64}
{\cal{S}}_{\alpha\beta}'=3(\zeta_{e(\alpha)}'-\zeta_{e(\beta)}').
\end{equation}
By using Eq.(\ref{e57}) and Eq.(\ref{e62}), we can obtain the evolution of the entropy perturbation
\begin{widetext}
\begin{equation}\label{65}
{\cal{S}}_{\alpha\beta}'={\cal{A}}_{intr(\alpha\beta)}+{\cal{B}}_{iso(\alpha\beta)}+{\cal{C}}_{quan(\alpha\beta)},
\end{equation}
where
\begin{eqnarray}
{\cal{A}}_{intr(\alpha\beta)}&=&3(1-f+f_1){\cal{H}}\left(\frac{3{\cal{H}}(1-f+f_1)\delta P_{intr(\alpha)}-\delta Q_{intr(\alpha)}}{\rho_{e(\alpha)}'}
-\frac{3{\cal{H}}(1-f+f_1)\delta P_{intr(\beta)}-\delta Q_{intr(\beta)}}{\rho_{e(\beta)}'}\right),\\
{\cal{B}}_{iso(\alpha\beta)}&=&-\sum_{\gamma}\frac{\rho_{e(\gamma)}'}{2\rho_e}\left(\frac{Q_{\alpha}}{\rho_{e(\alpha)}'}{\cal{S}}_{\alpha\gamma}
-\frac{Q_{\beta}}{\rho_{e(\beta)}'}{\cal{S}}_{\beta\gamma}\right),\\
{\cal{C}}_{quan(\alpha\beta)}&=&-3(1-f+f_1){\cal{H}}\left(\frac{Q_{\alpha}}{\rho_{e(\alpha)}'}\mathcal{E}_{\alpha}
-\frac{Q_{\beta}}{\rho_{e(\beta)}'}\mathcal{E}_{\beta}\right)\\
&=&-3(2f-f_1){\cal{H}}\left[\frac{1}{3}\left(\frac{Q_{\alpha}}{\rho_{e(\alpha)}'}\sum_{\gamma}\frac{\rho_{e(\gamma)}'}{\rho_e'}{\cal{S}}_{\alpha\gamma}
-\frac{Q_{\beta}}{\rho_{e(\beta)}'}\sum_{\gamma}\frac{\rho_{e(\gamma)}'}{\rho_e'}{\cal{S}}_{\beta\gamma}\right)
+\zeta_e\left(\frac{Q_{\alpha}}{\rho_{e(\alpha)}'}-\frac{Q_{\beta}}{\rho_{e(\beta)}'}\right)
\right.\nonumber\\
&&+\left.(\sigma+1){\cal{H}}\left(\frac{Q_{\alpha}\delta\rho_{e(\alpha)}}{\rho_{e(\alpha)}'^2}-\frac{Q_{\beta}\delta\rho_{e(\beta)}}{\rho_{e(\beta)}'^2}\right)\right].\label{e67}
\end{eqnarray}
\end{widetext}
Where, the relationship
\begin{equation}\label{e66}
\zeta_{e(\alpha)}=\zeta_e+\frac{1}{3}\sum_{\gamma}\frac{\rho_{e(\gamma)}'}{\rho_e'}{\cal{S}}_{\alpha\gamma}
\end{equation}
is used in the last equation.

As we can see that, there are three sources for the entropy perturbation. The first is the intrinsic non-adiabatic perturbation ${\cal{A}}_{intr(\alpha\beta)}$, the second is the entropy perturbation of others fluid ${\cal{B}}_{iso(\alpha\beta)}$, and the third comes from the effective quantum correction ${\cal{C}}_{quan(\alpha\beta)}$.

There are also three parts in the quantum source. The first part comes from the entropy perturbation of others fluid. The second part comes from the adiabatic curvature perturbation $\zeta_e$. In general speaking, the third term of Eq.(\ref{e67}) can not be represented by ${\cal{S}}_{\alpha\beta}$ completely, so it depends on the adiabatic curvature perturbation, too.

So we come to a conclusion that  the non-adiabatic entropy on the large scales can be driven by the adiabatic curvature perturbation. This conclusion is different from the classical cosmology, and this adiabatic source for non-adiabatic perturbations is on the quantum order.

However, in some special conditions, the quantum adiabatic source will be vanish. From Eq.(\ref{e67}), we know that, if $Q_{\alpha}/\rho_{e(\alpha)}'$ is the same for all fluids, the second term of Eq.(\ref{e67}) is equal to zero, and the third term can be represented by the entropy perturbation ${\cal{S}}_{\alpha\beta}$. Under this condition, the evolution of the entropy perturbation only depends on the non-adiabatic perturbation, and itself obeys a homogeneous second-order equation on the super-Hubble scales. It is the same as the classical conclusion.

\section{\label{s6}Decay model}

After deriving a general formalism of the adiabatic and the non-adiabatic perturbations for loop quantum cosmology, this section we apply it to a simple decay model, that is, a specific case of the non-relativistic matter $\vartheta$ decaying into radiation $\varsigma$. This process could be used in the curvaton scenario \cite{cur1,cur2,cur3,cur4}. In the model considered here, we assume that the energy density is dominated by the radiation $\rho_{\varsigma}$ and it is unperturbed, i.e. $\delta\rho_{\varsigma}\simeq 0$.

 We assume that the matter $\vartheta$ is unstable. It can decay into a radiation $\varsigma$ with a decay rate $\Xi$.  In our discussion, the decay rate $\Xi$ is treated as a constant, and the energy transfers from the pressureless fluid to the radiation fluid. We will give the evolving equations for the adiabatic and the non-adiabatic perturbations and solve them numerically. At the same time, we will give the results of the classical perturbations theory, used as a comparison.

\subsection{Background}

The energy transferring from $\vartheta$ to $\varsigma$ is described by
\begin{equation}\label{e70}
Q_{\vartheta}=-Q_{\varsigma}=-\Xi\rho_{\vartheta}.
\end{equation}
And the energy conservation equations are
\begin{eqnarray}
\rho'_{\vartheta}&=&-\rho_{\vartheta}\left(3{\cal{H}}-{\cal{H}}\frac{d\ln\nu}{d\ln p}+\Xi\right),\label{e71}\\
\rho'_{\varsigma}&=&-4{\cal{H}}\left(1-\frac{1}{3}\frac{d\ln\nu}{d\ln p}\right)\rho_{\varsigma}+\Xi\rho_{\vartheta},\label{e72}\\
\rho_e'&=&-{\cal{H}}\left(1-\frac{1}{3}\frac{d\ln\nu}{d\ln p}\right)(3\rho_{\vartheta}+4\rho_{\varsigma}).\label{e73}
\end{eqnarray}
Also the Fridemann equation is
\begin{equation}\label{e74}
{\cal{H}}^2=\frac{8\pi G}{3}\alpha p(\rho_{\vartheta}+\rho_{\varsigma}).
\end{equation}

It is convenient to introduce the dimensionless density parameters $\Omega_{\vartheta}, \Omega_{\varsigma}$
and the reduced decay rate $g$:
\begin{equation}\label{e75}
\Omega_{\vartheta}\equiv\frac{\rho_{\vartheta}}{\rho_e},\ \ \Omega_{\varsigma}\equiv\frac{\rho_{\varsigma}}{\rho_e},\ \ g\equiv\frac{\Xi}{\Xi+{\cal{H}}}.
\end{equation}
After that, the Eqs.(\ref{e71})-(\ref{e74}) can be rewritten as:
\begin{equation}
\partial _{\ln a}\Omega _\vartheta =\Omega _\vartheta \left[ \left( 1-\frac 1%
3\frac{d\ln \nu }{d\ln p}\right) \Omega _\varsigma -\frac g{1-g}\right] ,
\label{e76}
\end{equation}
\begin{equation}
\partial _{\ln a}\Omega _\varsigma =\Omega _\vartheta \left[ \frac g{1-g}%
-\left( 1-\frac 13\frac{d\ln \nu }{d\ln p}\right) \Omega _\varsigma \right] ,
\label{e77}
\end{equation}
\begin{eqnarray}
\partial _{\ln a}g &=&\frac{g(1-g)}2\left[ \left( 1-\frac 13\frac{d\ln \nu }{%
d\ln p}\right) (4-\Omega _\vartheta )\right.   \nonumber \\
&&\left. +1+\frac{d\ln \alpha }{d\ln p}\right]   \label{e78}
\end{eqnarray}

From the definitions of $\alpha$ and $\nu$ we note that:
\begin{equation}\label{e79}
\frac{d\ln\alpha}{d\ln p}\propto\frac{d\ln\nu}{d\ln p}\propto \exp(-\sigma\ln a).
\end{equation}
So the Eqs.(\ref{e76})-(\ref{e78}) can be seen as an autonomous system.

\begin{figure}
\includegraphics[width=0.5\textwidth]{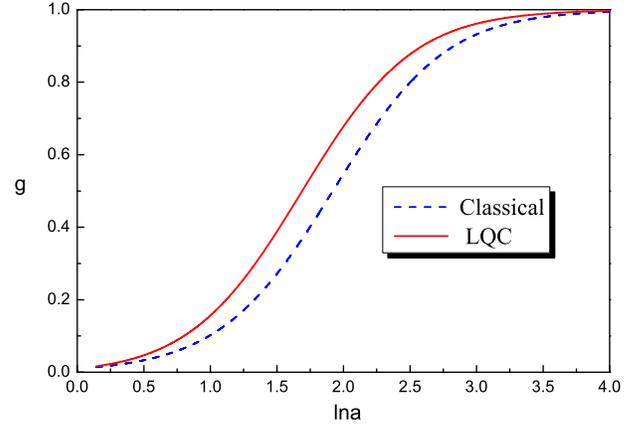}
\caption{The evolution of the reduced decay rate $g$ as a function of $\ln a$.
Here, the initial density $\Omega_{\vartheta}$, the initial decay rate $g$ and the parameter $\sigma$ are taken as, respectively, $\Omega_{\vartheta}=10^{-1}$, $g=10^{-2}$ and $\sigma=2$. The classical evolution is represented by the dashline.}\label{f1}
\end{figure}

\begin{figure}
\includegraphics[width=0.5\textwidth]{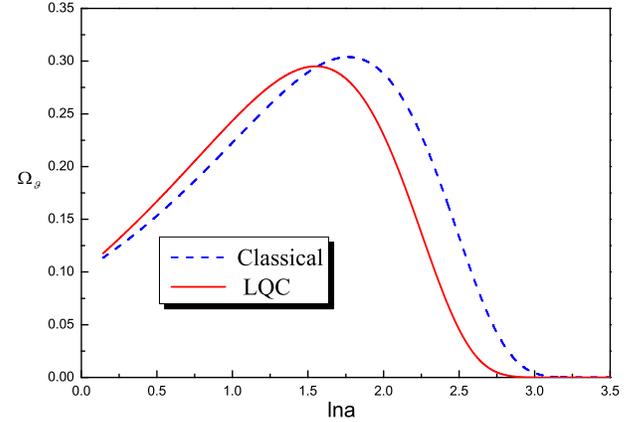}
\caption{The evolution of the dimensionless density parameter $\Omega_{\vartheta}$ as a function of $\ln a$.
Here, the initial density $\Omega_{\vartheta}$, the initial decay rate $g$ and the parameter $\sigma$ are taken as, respectively, $\Omega_{\vartheta}=10^{-1}$, $g=10^{-2}$ and $\sigma=2$. The classical evolution is represented by the dashline.}\label{f2}
\end{figure}

However, there is a constraint to this system:
\begin{equation}\label{e80}
\Omega_{\vartheta}+\Omega_{\varsigma}=1.
\end{equation}
Therefore, there are only two independent dynamical equations. The solutions with a fixed initial condition are illustrated in Fig.\ref{f1} and Fig.\ref{f2}.
We can see that, the expansion of universe and the decay of matter $\vartheta$ are both faster than classical evolution.


\subsection{Perturbations}
Both $\vartheta$ and $\varsigma$ have fixed equations of state, hence they meet the intrinsic adiabatic condition,
i.e. $\delta P_{intr(\vartheta)}=\delta P_{intr(\varsigma)}=0$. However, there is a nonzero entropy perturbation ${\cal{S}}_{\vartheta\varsigma}$,
so the curvature pertubation $\zeta_e$ is not a conserved quantity on the large scales. From Eq.(\ref{e37}) and Eq.(\ref{e56}) we have:
\begin{equation}\label{e81}
\zeta_e'=\frac{(1+f-f_1){\cal{H}}}{3\rho_e'^2}\rho_{\vartheta}'\rho_{\varsigma}'{\cal{S}}_{\vartheta\varsigma}.
\end{equation}
The perturbed energy transferring is given by
\begin{equation}\label{e82}
\delta Q_{\vartheta}=-\delta Q_{\varsigma}=-\Xi\delta\rho_{\vartheta},
\end{equation}
where we assume $\Xi$ is fixed by microphysics, i.e. $\delta\Xi=0$.

The energy transferring of $\vartheta$ is determined only by its energy density, therefore $\delta Q_{intr(\vartheta)}=0$.
However, for radiation $\varsigma$, its energy transferring depends on the decay of $\vartheta$, so $\delta Q_{intr(\varsigma)}\neq0$.
We can find that
\begin{equation}\label{e83}
\delta Q_{intr(\varsigma)}=\delta Q_{\varsigma}-\frac{Q_{\varsigma}'}{\rho_{\varsigma}'}\delta\rho_{\varsigma}=-\frac{\Xi\rho_{\vartheta}'}{3{\cal{H}}(1-f+f_1)}
{\cal{S}}_{\vartheta\varsigma}.
\end{equation}
Under our hypothesis, $\delta\rho_{\varsigma}\simeq0$, we have
\begin{equation}\label{e84}
{\cal{S}}_{\vartheta\varsigma}\simeq-3(1-f+f_1){\cal{H}}\frac{\delta\rho_{\vartheta}'}{\rho_{\vartheta}'}.
\end{equation}
Therefore, we can obtain the evolution equation for the entropy perturbation:
\begin{equation}\label{e86}
{\cal{S}}_{\vartheta\varsigma}'={\cal{A}}_{intr(\vartheta\varsigma)}+{\cal{B}}_{iso(\vartheta\varsigma)}+{\cal{C}}_{quan(\vartheta\varsigma)},
\end{equation}
where
\begin{eqnarray}
{\cal{A}}_{intr(\vartheta\varsigma)}&=&-\Xi\frac{\rho_{\vartheta}'}{\rho_{\varsigma}'}{\cal{S}}_{\vartheta\varsigma},\\
{\cal{B}}_{iso(\vartheta\varsigma)}&=&\frac{\Xi\rho_{\vartheta}}{2\rho_e}\left(\frac{\rho_{\varsigma}'}{\rho_{\vartheta}'}-\frac{\rho_{\varsigma}'}{\rho_{\vartheta}'}\right){\cal{S}}_{\vartheta\varsigma},\\
{\cal{C}}_{quan(\vartheta\varsigma)}&=&-3{\cal{H}}(2f-f_1)\Xi\rho_{\vartheta}\left(\frac{1}{\rho_{\vartheta}'}+\frac{1}{\rho_{\varsigma}'}\right)\zeta_e\nonumber\\
&&-{\cal{H}}(2f-f_1)\Xi\rho_{\vartheta}\left[\frac{\rho_{\vartheta}'}{\rho_{\varsigma}'\rho_e'}-\frac{\rho_{\varsigma}'}{\rho_{\vartheta}'\rho_e'}\right.\nonumber\\
&&\left.+(\sigma+1)\frac{1}{\rho_{\vartheta}'}\right]{\cal{S}}_{\vartheta\varsigma}.\label{e87}
\end{eqnarray}

Eq.(\ref{e81}) and Eq.(\ref{e86}) form a closed system of the first-order equations for the evolution of the adiabatic perturbation $\zeta_e$ and the entropy perturbation ${\cal{S}}_{\vartheta\varsigma}$ on the large scales.

\begin{figure}
\includegraphics[width=0.5\textwidth]{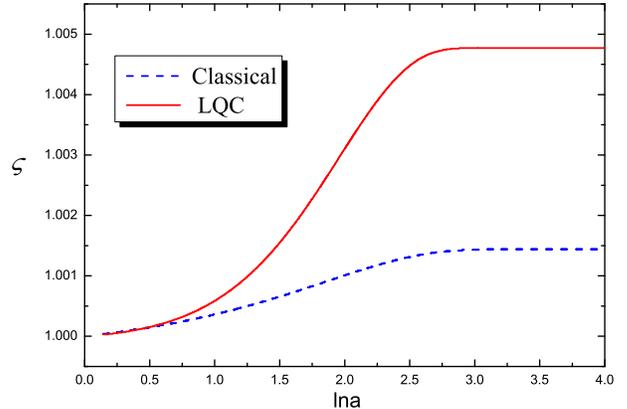}
\caption{The evolution of the adiabatic perturbation $\zeta$ as a function of $\ln a$.
Here, the initial values of the related parameters are taken as:
the initial density $\Omega_{\vartheta}=10^{-1}$, the initial decay rate $g=10^{-2}$, the initial adiabatic perturbation $\zeta=1$, the initial entropy perturbation ${\cal{S}}_{\vartheta\varsigma}=10^{-2}$ and the parameter $\sigma=2$.  The classical evolution is represented by the dashline.}
\label{f3}
\end{figure}

The following discussion, we focus on the evolution of the $\zeta_e$. The numerical result is shown in Fig.\ref{f3}.
In our model, the initial entropy perturbation is positive. From the Eq.(\ref{e81}) we know that $\zeta_e$ should increase as the universe expanding. However, as $\vartheta$ decaying into $\varsigma$, the entropy perturbation must be vanish, and the $\zeta_e$ becomes a conserved quantity on the large scales. From Fig.\ref{f3} we can see that, the final $\zeta_e$ is bigger than the classical one. In the classical theory, the evolution of the entropy perturbation is independent, the evolution of $\zeta$ does not impact on the entropy perturbation. On the other hand, in LQC, the evolution of the entropy is affected by $\zeta_e$, and this effect is impact on the evolution of $\zeta_e$ itself. So we can see the different final value of $\zeta$ in Fig.\ref{f3}.

\section{\label{s5}Summary and conclusions}

What we study in this paper are the adiabatic and the non-adiabatic cosmological perturbations with the inverse triad correction of LQC.

In order to discuss the general universe model, we need a general form of the perturbations theory of LQC. The complete theory should come from the analysis of the effective Hamiltonian like in \cite{lqcg1,lqcg2}. However, the gauge-invariant form of this theory has yet to be addressed. So, we have to rewrite the perturbations theory of LQC to be a hydrodynamical formalism. In principle, this formalism can only be applied to the universe fulfilled by scalar field, and take the scalar field as a special fluid. But in this paper, we assume this effective form can also be applied to a general fluid, and we give the definition of the effective curvature perturbation on an uniform density hypersurfaces $\zeta_e$.

In the classical theory for the cosmological perturbations, the curvature perturbation on the uniform density hypersurfaces $\zeta$ is gauge-invariant and conservative on the large scales. So in our definition of the effective curvature perturbation, it is required to have similar properties. The requirements of the gauge-invariant and the conservation both lead to one of the counterterms $g_1=0$, i.e. Eq.(\ref{e45}). Therefore, Eq.(\ref{e45}) can be seen as a restriction to the space of ambiguity parameters $(\sigma, q, \ell)$.

In a further discussion, we generalize the hydrodynamical form of theory to a multi-fluids model. There are interaction between the different fluids in this model. So there will be a non-adiabatic entropy perturbation which reflects the difference of the curvature perturbation between the different fluids.

In the classical theory, the entropy perturbation can evolve into an adiabatic curvature perturbation at late time on the large scales, and itself can evolve independently. In other words, there is no any adiabatic source for the non-adiabatic entropy perturbation. However, we find that, in the effective theory of LQC, there will be a quantum adiabatic source for the non-adiabatic entropy perturbation. In a more general model of the universe, this source does not disappear, except for some special situation in which $Q_{\alpha}/\rho_{e(\alpha)}'$ is the same for all fluids. From this we know that, the reason of the emergence of the quantum adiabatic source is an asynchronous change of $Q_{\alpha}/\rho_{e(\alpha)}'$. This is similar to the entropy perturbation which comes from the asynchronous change of $\delta\rho_{e(\alpha)}/\rho_{e(\alpha)}'$. And we apply this effective formalism to a simple decay model in which a nonrelativistic matter $\vartheta$ decaying into a radiation $\varsigma$. We find that, the final value of $\zeta_e$ is bigger than its counterparty in the classical theory.

\begin{acknowledgments}

This work was supported by the National Natural Science Foundation of China under Grant Nos. 10875012, 11175019 and the Fundamental Research Funds for the Central Universities.
\end{acknowledgments}

\end{document}